# Uniform Patterns of Fe Vacancy Ordering in the $K_x(Fe,Co)_{2-y}Se_2$ Superconductors

*Sergey M. Kazakov[1*], Artem M. Abakumov,[2] Santiago González[3], Juan Manuel Perez-Mato[3], Alexander V. Ovchinnikov[1], Maria V. Roslova[1], Alexander I. Boltalin[1], Igor V. Morozov[1], Evgeny V. Antipov[1], Gustaaf Van Tendeloo[2]*

[1] Department of Chemistry, Moscow State University, 119991 Moscow, Russia

[2] *Electron Microscopy for Materials Research (EMAT), University of Antwerp, Groenenborgerlaan 171, B-2020, Antwerp, Belgium*

[3] Dpto. de Física de la Materia Condensada, Facultad de Ciencia y Tecnología, Universidad del País Vasco, Apdo 644, Bilbao 48080, Spain

* Corresponding author, e-mail: kazakov@icr.chem.msu.ru

**Abstract**

The Fe-vacancy ordering patterns in the superconducting $K_xFe_{2-y}Se_2$ and non-superconducting $K_x(Fe,Co)_{2-y}Se_2$ samples have been investigated by electron diffraction and high angle annular dark field scanning transmission electron microscopy. The Fe-vacancy ordering occurs in the *ab* plane of the parent $ThCr_2Si_2$-type structure, demonstrating two types of patterns. The superstructure I retains the tetragonal symmetry and can be described with the $a_I = b_I = a_s\sqrt{5}$ ($a_s$ is the unit cell parameter of the parent $ThCr_2Si_2$-type structure) supercell and $I4/m$ space group. The superstructure II reduces the symmetry to orthorhombic with the $a_{II} = a_s\sqrt{2}$, $b_{II} = 2a_s\sqrt{2}$ supercell and the *Ibam* space group. This type of superstructure is observed for the first time in $K_xFe_{2-y}Se_2$. The Fe-vacancy ordering is inhomogeneous: the disordered areas interleave with the superstructures I and II in the same crystallite. The observed superstructures represent the compositionally-dependent uniform ordering patterns of two species (the Fe atoms and vacancies) on a square lattice. More complex uniform ordered configurations, including compositional stripes, can be predicted for different chemical compositions of the $K_xFe_{2-y}Se_2$ (0 < y < 0.5) solid solutions.

**Introduction**

The discovery of high-temperature superconductivity in La(O,F)FeAs has triggered a burst of interest in new compounds bearing *anti*-fluorite ($Fe_2As_2$) or ($Fe_2Ch_2$) (Ch = S, Se, Te) layers.[1] The tetragonal FeSe phase was found to be superconducting with $T_c$ of 8 K and 37 K under pressure of 7 GPa.[2,3] The crystal structure of FeSe is the simplest among the Fe-based superconductors; it consists of only the ($Fe_2Se_2$) layers with the PbO-type structure. Intercalation of alkali metals into this material lead to the discovery of superconducting $A_xFe_{2-y}Se_2$ (A=K, Rb, Cs, Tl) phases with $T_c$ above 30 K.[4-7] These compounds can be tentatively described as derivatives of the tetragonal $ThCr_2Si_2$ structure (also called the 122-type structure, Fig. 1). The ($Fe_2Se_2$) layers can adopt some Fe deficiency and the ordering of the Fe atoms and vacancies arises at sufficient concentration of the vacant Fe sites. Different types of the Fe-vacancy ordering in $Tl_yFe_{2-x}Ch_2$ (Ch = S, Se) were observed long time ago.[8,9] Due to renewed interest to the superconducting iron pnictides and chalcogenides many new experimental studies on the vacancy ordering in $A_yFe_{2-x}Se_2$ (X=K, Tl, Cs) have been performed very recently.[10-15] Superlattice peaks corresponding a (√5, √5, 1) superstructure have been observed in a transmission electron microscopy study of superconducting $K_{0.8}Fe_{2-y}Se_2$ (0.2 ≤y≤0.5) samples, and they were attributed to Fe-vacancy ordering.[16] The same superstructure was found in single crystal X-ray diffraction studies on similar superconducting[11] or non-superconducting[12] samples; the origin of the superstructure was defined as ordering of the Fe atoms and vacancies in the *ab* plane of the parent $ThCr_2Si_2$ structure. Neutron diffraction studies also demonstrate an ordered vacancy structure with the ($\sqrt{5}$, $\sqrt{5}$, 1) supercell.[10,14,15]

The insulating and the superconducting state are both observed in $K_xFe_{2-y}Se_2$ depending on the composition.[7,17,18] It is assumed that the interplay between the Fe vacancy ordering patterns is crucial for the occurrence of superconductivity in this system. Neutron diffraction study underlined the fact that superconductivity only occurs in the $K_xFe_{2-y}Se_2$ when the Fe content is compatible with the vacancy ordering pattern.[10] However, Han *et al* conclude that the superconductivity is achieved when the Fe-vacancies are in a random (disordered) state.[19] In this contribution we report on the local scale Fe-vacancy ordering in superconducting $K_xFe_{2-y}Se_2$ and non-superconducting $K_x(Fe,Co)_{2-y}Se_2$ samples. Transmission electron microscopy investigation of both samples did not reveal noticeable difference in their microstructure, demonstrating inhomogeneous and incomplete ordering of the Fe atoms and vacancies and a presence of different ordering patterns, which allow us to predict more complex ordering combination, including compositional stripes.

**Experimental Section**

Single crystals of $K_x(Fe,Co)_{2-y}Se_2$ were grown from the melt by self flux method as described previously.[17] K pieces and FeSe (CoSe) powders with nominal compositions $K_{0.8}Fe_2Se_2$ and $K_{0.8}(Fe_{0.975}Co_{0.025})_2Se_2$ were loaded in alumina crucibles in an Ar-filled glove box, evacuated and sealed in double walled quartz ampoules. The ampoules were heated simultaneously in the same furnace to 1090°C at 100°C/ h, held at this temperature for 5 h and then were cooled to 800°C during 54 h after which ampoules were taken out from the furnace.

The phase composition of the samples was checked with powder X-ray diffraction using a Huber G670 Guinier diffractometer ($CuK_{\alpha 1}$-radiation, curved Ge monochromator, transmission mode, image plate). The chemical composition of the crystals was verified with energy-dispersive X-ray microanalysis performed with a JEOL JSM 6490LV scanning electron microscope (SEM) equipped with an Oxford Instruments attachment. The chemical compositions of the crystals correspond to the $K_{0.84(3)}Fe_{1.78(3)}Se_2$ and $K_{0.84(3)}Fe_{1.77(4)}Co_{0.03(1)}Se_2$ formulae. According to the magnetic susceptibility measurements, the $K_xFe_{2-y}Se_2$ sample demonstrates superconducting transition with $T_c$ = 17K, whereas the $K_x(Fe,Co)_{2-y}Se_2$ sample is non-superconducting.

The specimens for transmission electron microscopy investigation (TEM) were prepared by crushing the $K_x(Fe,Co)_{2-y}Se_2$ single crystals in hexane and depositing the suspension on a holey carbon grid. The specimen were prepared in an Ar-filled glove box and transported into TEM under Ar. Electron diffraction (ED) patterns and high angle annular dark field scanning TEM (HAADF-STEM) images were recorded on a Titan G3 80-300 microscope equipped with the probe aberration corrector and operated at 120 kV. High resolution TEM (HRTEM) images were taken with a Tecnai G2 microscope operated at 200 kV. Observations at the accelerating voltages of 300 kV, 200 kV and 120 kV revealed that, although the parent $ThCr_2Si_2$-type structure is not noticeably affected by the electron beam, the superstructure associated with the Fe-vacancy ordering suffers from irradiation damage. The beam damage appears as gradual decrease of intensity of the superlattice diffraction spots originating from the Fe-vacancy ordering up to their complete disappearance. This is accompanied by suppressing of characteristic contrast on the HAADF-STEM images indicating a formation of disordered structure. This transformation is irreversible. The superstructure is moderately stable at 200 kV making the HRTEM observations possible. Observations of the Fe-vacancy ordering in the HAADF-STEM regime were possible only at 120 kV. Because of low signal-to-noise ratio, some of the HAADF-STEM images were Fourier filtered.

**Results**

The electron diffraction (ED) patterns of the $K_x(Fe,Co)_{2-y}Se_2$ samples appear to be rather complex (Fig. 2). The brightest spots belong to the tetragonal body-centered $ThCr_2Si_2$-type subcell with $a_s = b_s \approx 3.9$Å, $c_s \approx 14.1$Å. Weaker spots are the supercell reflections. The ED in Fig. 2a was taken from a relatively large area of the crystallite. Careful analysis with smaller selected area aperture revealed that it contains at least three types of superstructure, schematically shown in Fig. 2d. The first set of superstructure reflections (marked as open squares in Fig. 2d) can be indexed on a tetragonal body-centered supercell with the unit cell parameters $a_I = b_I = a_s\sqrt{5} \approx 8.7$Å (superstructure **I**). The lattice vectors of the supercell **I** in $ab$ plane are related to those of the subcell as $\mathbf{a}_I = 2\mathbf{a}_s + \mathbf{b}_s$, $\mathbf{b}_I = \mathbf{a}_s - 2\mathbf{b}_s$. The positions of the superlattice reflections belonging to the second set (marked as filled squares in Fig. 2d) also demonstrate apparent tetragonal symmetry (superstructure **II**). However, more careful examination of the reflections related by a 90° rotation revealed that their intensities are clearly different. This indicates that the superstructure **II** results in decreasing symmetry down to orthorhombic and is present in two twinned variants. The unit cell of this superstructure is based on face diagonals of the parent $ThCr_2Si_2$-type subcell with doubling of the repeat period along one direction: $a_{II} = a_s\sqrt{2} \approx 5.5$Å, $b_{II} = 2a_s\sqrt{2} \approx 11$Å ($\mathbf{a}_{II} = \mathbf{a}_s + \mathbf{b}_s$, $\mathbf{b}_{II} = 2\mathbf{a}_s - 2\mathbf{b}_s$). The superlattice reflections of the third set are positioned at $h+1/2, k+1/2, 0$ along both [110] and [-110] directions (superstructure **III**, marked as asterisks in Fig. 2d). Different intensities of the superlattice reflections along these directions and absence of the $h00$, $h$ = odd and $0k0$, $k$ = odd reflections point towards 90° rotational twins of two orthorhombic variants with $a_{III} \approx b_{III} = a_s\sqrt{2} \approx 5.5$Å ($\mathbf{a}_{III} = \mathbf{a}_s + \mathbf{b}_s$, $\mathbf{b}_{III} = \mathbf{a}_s - \mathbf{b}_s$). The superstructures **I** and **II** can be observed in separate parts of the crystallites. The sets of the superlattice reflections corresponding to the superstructures **I**+**III** and **II**+**III** are shown in Fig. 2b and 2c, respectively. Note that Fig. 2c clearly demonstrates the orthorhombic symmetry for the superstructure **II**.

HAADF-STEM observations revealed that the superstructures **I** and **II** originate from ordering of the Fe atoms and vacancies. On the [001] HAADF-STEM image of the area with the ordering of type **I** (Fig. 3) the rectangular mesh of the bright dots corresponds to the projections of the mixed K and Se columns. The Fe columns center the squares of the bright dots. The darker spot at the center of the bright dot square correspond to lower scattering density due to presence of the Fe vacancies. Fourier transform of the image clearly demonstrates the reflections corresponding to the type **I** superstructure (Fig. 3). Note that the spots of the superstructure **III** are absent on the Fourier transform. It reflects that the superstructures **I** and **III** are of different nature and do not necessarily co-exist with each other. The Fe-vacancy ordering is not regular: the areas where the arrangement of the dark spots obeys the supercell **I** co-exist with the areas where the Fe-vacancy ordering is either locally violated (as two adjacent vacant Fe sites marked with arrowheads in Fig. 3) or apparently absent (the area at the left bottom corner in Fig. 3). Such local disorder is in agreement with the single crystal structure data on $K_{0.86}Fe_{1.56}Se_2$ demonstrating incomplete ordering of the Fe atoms and vacancies with the occupancy factors of 0.920 and 0.227 for the Fe-rich and Fe-depleted sites, respectively.[12] The ordered areas are present as patches with the size of few tens of nm. The enlarged image of such local area with the type **I** superstructure is shown in Fig. 4. Vacant Fe columns are placed at the 0,0,z and 1/2,1/2,z positions of the $a_I = b_I = a_s\sqrt{5}$ supercell (outlined in Fig. 4). Such ordering of the Fe vacancies is incompatible with the mirror planes parallel to the *c*-axis in the *I*4/*mmm* space group of the parent $ThCr_2Si_2$ structure and reduces the symmetry down to *I*4/*m*. This symmetry reduction is obvious also from the ED pattern in Fig. 2b and the Fourier transform in Fig. 3.

The Fe-vacancy ordering pattern corresponding to the type **II** superstructure is shown in Fig. 5. The vacant Fe columns are also placed at the 0,0,z and 1/2,1/2,z positions of the $a_{II} = a_s\sqrt{2} \approx 5.5$Å, $b_{II} = 2a_s\sqrt{2}$ supercell. The maximal possible symmetry of such ordering was analyzed with the FINDSYM program[20] assuming that the superstructure is caused exclusively by alternation of the vacant and occupied Fe sites. The space group corresponding to the type **II** superstructure is *Ibam*, but the actual symmetry might correspond to one of its subgroups due to possible atomic displacements, not visible on the HAADF-STEM image. The dots corresponding to the projected Fe-depleted columns in the HAADF-STEM image of the superstructure **II** are generally weaker than the dots of the Fe-rich columns, but their intensity often does not vanish completely (Fig. 5). This indicates that some scattering density remains in the Fe-depleted columns and that the Fe-vacancy ordering is incomplete.

The superstructure **III** does not clearly show up on the HAADF-STEM images. The superstructure **III** is clearly visible only on the [001] HRTEM image as alternating rows of dots with clearly different brightness separated by ~ 5.5Å (Fig. 6). The image confirms the orthorhombic symmetry associated with this type of superstructure. Although the image does not provide sufficient information to reveal possible origin of the type **III** superstructure one can speculate that this type superstructure is not caused by the Fe-vacancy ordering but rather by K-ion ordering.

**Discussion**

TEM investigation revealed that the ordering of the Fe atoms and vacancies is present in the structures of the superconducting $K_xFe_{2-y}Se_2$ and non-superconducting $K_x(Fe,Co)_{2-y}Se_2$ samples. This ordering is intrinsically inhomogeneous at the local scale: small patches of the ordered areas co-exist with the areas without visible ordering of the Fe atoms and vacancies; different ordering patterns are present in the same crystallite. Some residual occupation of the Fe-depleted columns by the Fe atoms remains, resulting in incomplete ordering even at the areas where regular superstructure is observed. These effects are responsible for deviations between the ideal chemical compositions corresponding to the observed superstructures and the nominal composition of the samples. The fact that the similar microstructure was observed in both superconducting $K_xFe_{2-y}Se_2$ and non-superconducting $K_x(Fe,Co)_{2-y}Se_2$ samples may indicate that low level Co-doping does not influence the structural ordering though it leads to the extremely fast $T_c$ decrease. This behavior is very different from that of $AeFe_2As_2$ (Ae=Ba, Sr) compounds where the Co for Fe substitution results in inducing superconductivity in parent non-superconducting material.[21]

The Fe-vacancy ordering demonstrates two types of patterns. One (superstructure **I**) retains the tetragonal symmetry and can be described with the $a_I = b_I = a_s\sqrt{5}$ supercell and *I4/m* space group. The second one (superstructure **II**) reduces the symmetry to orthorhombic with the $a_{II} = a_s\sqrt{2}$, $b_{II} = 2a_s\sqrt{2}$ supercell and the *Ibam* space group. The Fe-vacancy ordering pattern with the same supercell and space group was observed before for the $TlFe_{2-y}S_2$ (y = 0.5) compound [8]. The schemes of the Fe-vacancy ordering patterns in the *ab* plane of both supercells are shown in Fig. 7. Assuming the idealized case of a complete ordering of the Fe atoms and vacancies, the superstructures **I** and **II** should have different chemical compositions corresponding to y = 0.4 and y = 0.5, respectively, in the $K_x(Fe,Co)_{2-y}Se_2$ formula. Thus, the co-existence of two types of superstructures can be attributed to local chemical inhomogeneity.

The observed ordering patterns can be discussed in terms of the compositionally-dependent ordering of two species on the $A_{1-\delta}B_\delta$ square lattice (A = Fe, B = vacancy, $\delta$ = y/2 in the $K_xFe_{2-y}Se_2$ formula). Both **I** (y = 2/5, $\delta$ = 1/5) and **II** (y = 1/2, $\delta$ = 1/4) vacancy ordering patterns can be understood as most *uniform* orderings compatible with the underlying square Fe lattice and associated with certain compositions. They can be satisfactorily explained just assuming that the Fe vacancies on the 2D square lattice are distributed as uniformly as possible within the plane.[22] In fact, the vacancy patterns **I** and **II** are identical to the ordering patterns on a $A_{1-\delta}B_\delta$ square lattice for $\delta$ = 1/5 (Fig. 8a) and $\delta$ = 1/4 (Fig. 8b) obtained by energy minimization of a 2D model where only a simple short range effective pair repulsive potential of Yukawa-type [23] between the B species (in our case - the Fe vacancies) is introduced. The patterns obtained in this way can be considered as uniform or pseudo-uniform. In most cases, these uniform orderings can be rationalized as those with the minority motifs having smaller average number of neighbors of the same type at the shortest relevant distance than any other one, regardless of the distribution of neighbors at longer distances. It may seem, for instance, surprising that the orthorhombic ordering pattern of the type **II** superstructure at $\delta$ = 1/4 (Fig. 8b) prevails over the more symmetric tetragonal supercell arrangement with $a = b = a_s\sqrt{2}$ (Fig. 8c). However, the orthorhombic pattern of the superstructure **II** reduces the number of neighboring Fe vacancies at the shortest distance of $2d_{110}$ of the basic structure from 4 to 2 in comparison with the tetragonal ordering of Fig. 8c. This reduction of the number of nearest neighbors is expected to be in general sufficient for making it more favorable as long as the vacancy ordering is driven by some general tendency to minimize the contacts between them.

The fact that the superstructures **I** and **II** follow such universal uniform ordering patterns allows predicting the most probable Fe-vacancy ordering patterns for other compositions of the $K_xFe_{2-y}Se_2$ (0 < y < 0.5) solid solutions, assuming that they will also be uniform 2D arrangements. Some specific compositions should result in oblique ordering patterns, such as the one expected in the case of y = 1/3 ($\delta$ = 1/6) with only monoclinic symmetry (Fig. 8d). The most interesting cases are however those where the most uniform patterns are associated with the formation of compositional stripes. For y = 2/7 ≈ 0.286 ($\delta$ = 1/7) the most uniform ordering pattern (Fig. 8e) is obtained by alternating single stripes of the uniform patterns corresponding to higher ($\delta$ = 1/6) and lower ($\delta$ = 1/8) densities of the Fe vacancies [21]. Following such scheme of concatenation of the basic uniform patterns, one can obtain the most uniform structure for y = 3/7 ≈ 0.429 ($\delta$ = 3/14) as a combination of one $\delta$ = 1/4 (superstructure **II**) stripe with two $\delta$ = 1/5 (superstructure **I**) stripes (Fig. 8f). Because such patterns will be strongly compositionally dependent, very careful control over the chemical composition and completeness of the Fe-vacancy ordering will be absolutely necessary to observe them experimentally. Although it might be very challenging to prepare the material with long range stripe cation ordering, such structures might be observed locally in the samples with systematic and controlled variation of the Fe vacancy concentration. Such work is currently in progress.

**Acknowledgement**

We thank Prof. A.N. Vasiliev for magnetic measurements. This work was partially supported by the Ministry of Science and Education of Russian Federation under the State contract P-279. The support of the Russian Foundation for Basic Research is acknowledged (Grant No. 10-03-00681-a) and RFBR-DFG (Project No. 10-03-91334).

**References**

1. Kamihara, Y.; Watanabe, T.; Hirano, M.; Hosono H. *J. Am. Chem. Soc*. **2008**, *130*, 3296.

2. Hsu, F-C.; Luo, J-Y.; Yeh, K-W.; Chen, T-K.; Huang, T-W.; Wu, P.M.; Lee, Y-C.; Huang, Y-L.; Chu, Y-Y.; Yan, D-C.; Wu, M-K. *Proc. Natl. Acad. Sci. U.S.A* **2008**, *105*, 14262.

3. Medvedev, S.; McQueen, T.M.; Troyan, I.A.; Palasyuk, T.; Eremets, M.I.; Cava, R.J.; Naghavi, S.; Casper, F.; Ksenofontov, V.; Wortmann, G.; Felser, C. *Nature Materials* **2009**, *8*, 630.

4. Guo, J.; Jin, S.; Wang, G.; Wang, S.; Zhu, K.; Zhou, T.; He, M.; Chen, X. *Phys. Rev. B* **2010**, *82*, 180520.

5. Wang, A.F.; Ying, J.J.; Yan, Y.J.; Liu, R.H.; Luo, X.G.; Li, Z.Y.; Wang, X.F.; Zhang, M.; Ye, G.J.; Cheng, P.; Xiang, Z.J.; Chen, X.H. *Phys. Rev. B* **2011**, *83*, 060512.

6. Krzton-Maziopa, A.; Shermadini, Z.; Pomjakushina, E.; Pomjakushin, V.; Bendele, V.M.; Amato, A.; Khasanov, R.; Luetkens, H.; Conder, K. *J. Phys.: Condens. Matter* **2011**, *23*, 052203.

7. Fang, M.; Wang, H.; Dong, C.; Li, Z.; Feng, C.; Chen, J.; Yuan, H.Q. *EPL* **2011**, *94*, 27009.

8. Zabel, M.; Range, K.J. *Rev. Chim. Miner.* **1980**, *17*, 561-568.

9. Haggstrom, L.; Seidel, A.; Berger, R. *J. Magn. Magn. Mater.* **1991**, *98*, 37.

10. Bao, W.; Li, G.N.; Huang, Q.; Chen, G.F.; He, J.B.; Green, M.A.; Qiu, Y.; Wang, D.M.; Luo, J.L. **2011**, arXiv:1102.3674.

11. Zavalij, P.; Bao, W.; Wang, X.F.; Ying, J.J.; Chen, X.H.; Wang, D.M.; He, J.B.; Wang, X.Q.; Chen, G.F.; Hsieh, P.; Huang, Q.; Green, M.A. *Phys. Rev. B.* **2011**, *83*, 132509.

12. Bacsa, J.; Ganin, A.Y.; Takabayashi, Y., Christensen, K.E.; Prassides, K.; Rosseinsky, M.J.; Claridge, J.B. *Chem. Sci.* **2011**, 2, 1054.

13. Ye, F.; Chi, S.; Bao, W.; Wang, X.F.; Ying, J.J.; Chen, X.H.; Wang, H.D.; Dong, C.H.; Fang. M. **2011**, arXiv:1102.2882.

14. Bao, W.; Huang, Q.; Chen, G.F.; Green, M.A.; Wang, D.M.; He, J.B.; Wang, X.Q.; Qiu, Y. *Chinese Phys. Lett.* **2011**, 28, 086104.

15. Pomjakushin, V.Yu.; Sheptyakov, D.V.; Pomjakushina, E.V.; Krzton-Maziopa, A.; Conder, K.; Chernyshov, D.; Svitlyk, V.; Shermadini, Z. *Phys. Rev. B* **2011**, *83*, 144410.

16. Wang, Z.; Song, Y.J.; Shi, H.L.; Wang, Z.W.; Chen, Z.; Tian, H.F.; Chen, G.F.; Guo, J.G.; Yang, H.X.; Li, J.Q. *Phys. Rev. B* **2011**, *83*, 140505(R).

17. Ying, J. J.; Wang, X. F.; Luo, X. G.; Wang, A. F.; Zhang, M.; Yan, Y. J.; Xiang, Z. J.; Liu, R. H.; Cheng, P.; Ye, G. J.; Chen X. H. *Phys. Rev. B* **2011**, *83*, 212502.

18. Luo, X.G.; Wang, X.F.; Ying, J.J.; Yan, Y.J.; Li, Z.Y.; Zhang, M.; Wang, A.F.; Cheng, P.; Xiang, Z.J.; Ye, G.J.; Liu, R.H.; Chen, X.H. *New J. Phys.* **2011**, *13*, 053011.

19. Han, F.; Shen, B.; Wang, Z-Y.; Wen, H-H. **2011**, arXiv:1103.1347.

20. Stokes, H.T.; Hatch, D.M. **2004**. FINDSYM, stokes.byu.edu/isotropy.html.

21. Leithe-Jasper, A.; Schnelle, W.; Geibel, C.; Rosner H. *Phys. Rev. Lett.* **2008**, *101*, 207004.

22. González, S.; Perez-Mato, J.M.; Elcoro, L.; García, A. *submitted for publication to Phys. Rev. B.* **2011**.

23. $V(r) = exp(-r/r_0)/r$, where the distance $r_0$ is in between the first and second shell of the B neighbors for a given density of the B species.

**Figures**

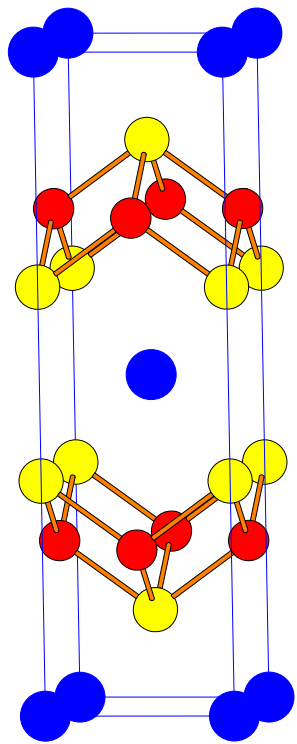

**Figure 1.** The parent ThCr$_2$Si$_2$-type *I*4/*mmm* structure of K$_x$Fe$_{2-y}$Se$_2$. The K, Fe and Se atoms are indicated as blue, red and yellow circles, respectively.

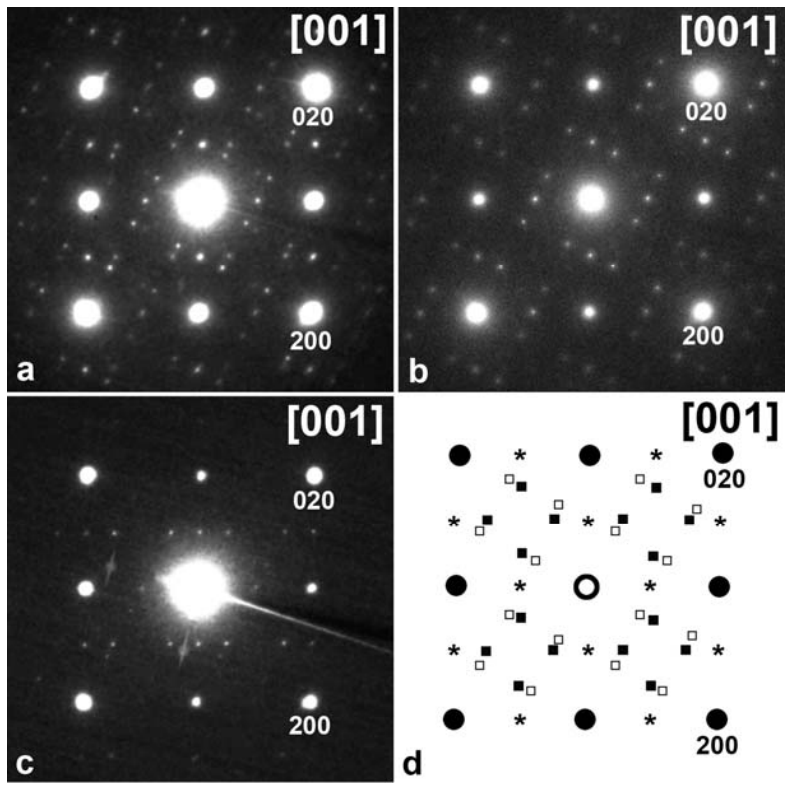

**Figure 2.** Electron diffraction patterns taken over large area of the crystallite (a) and with smaller selected area aperture (b, c). The scheme (d) represents the subcell reflections (black), and the reflections of the type **I**, **II** and **III** superstructures (open squares, filled squares and asterisks, respectively).

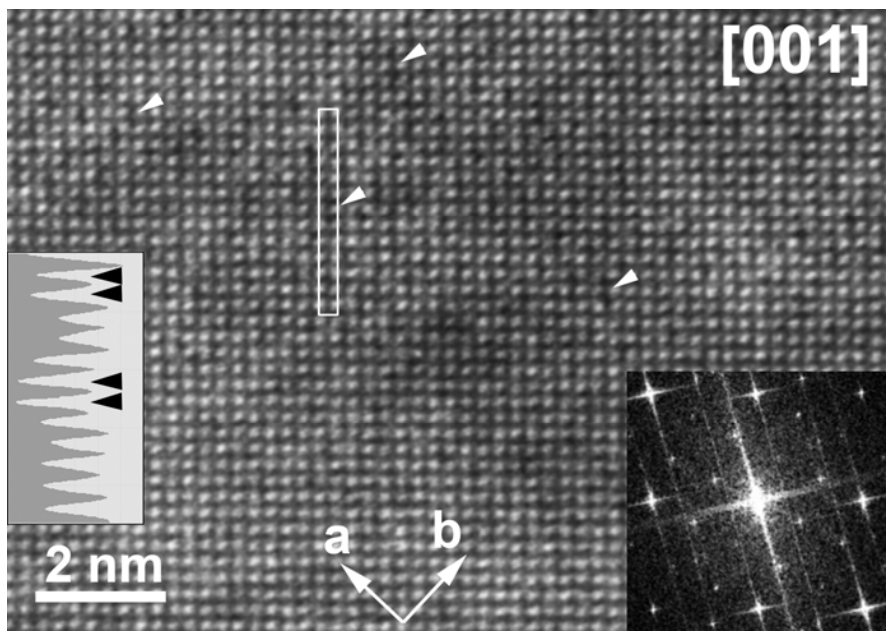

**Figure 3.** [001] HAADF-STEM image of the area with the type **I** superstructure. Fourier transform of the image is shown at right bottom corner. The adjacent pairs of the Fe-depleted columns are marked with arrowheads. The intensity variation produced by such pairs of adjacent columns is clearly seen in the intensity profile (inset at the left side, marked with black arrowheads) taken over the area marked with a white rectangle on the image.

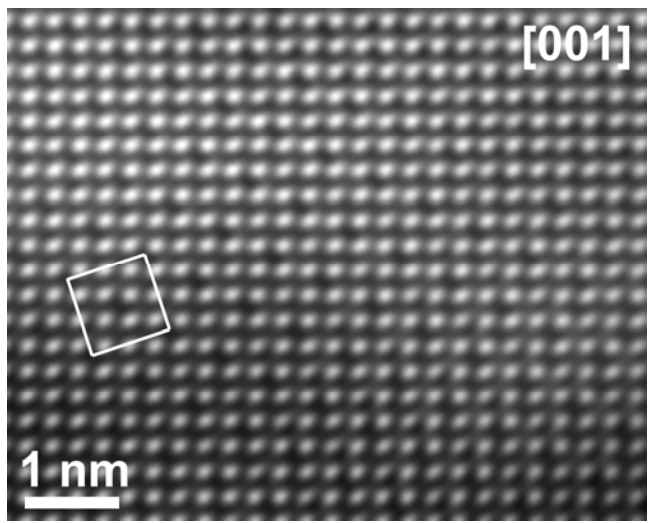

**Figure 4.** [001] HAADF-STEM image of the ordered area with the type **I** superstructure. The $a_I = b_I = a_s\sqrt{5}$ supercell is outlined.

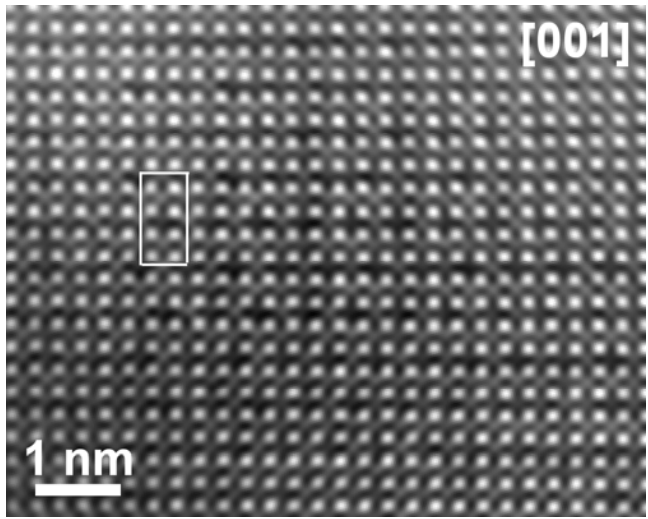

**Figure 5.** [001] HAADF-STEM image of the ordered area with the type **II** superstructure. The $a_{II} = a_s\sqrt{2}$, $b_{II} = 2a_s\sqrt{2}$ supercell is outlined.

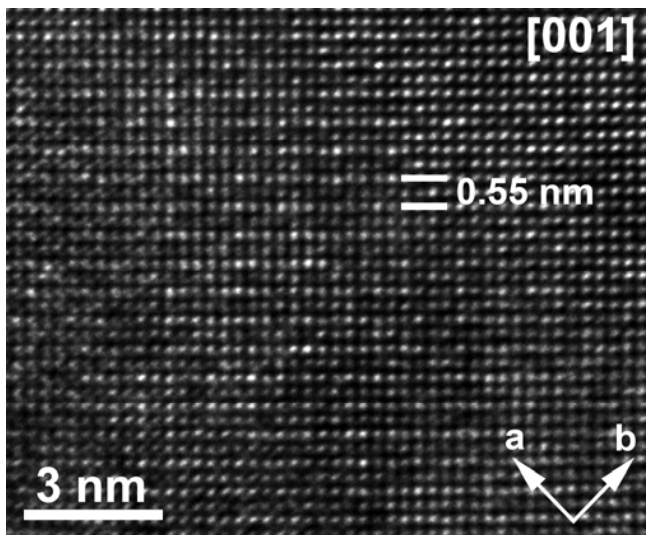

**Figure 6.** [001] HRTEM image of the ordered area with the type **III** superstructure.

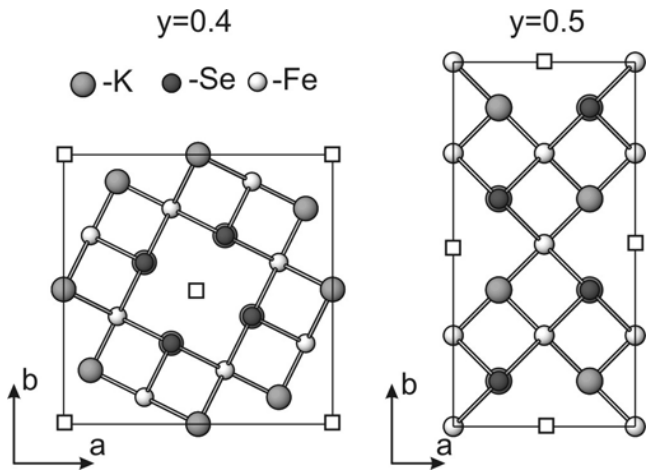

**Figure 7.** The schemes of the Fe-vacancy ordering patterns in $K_x(Fe,Co)_{2-y}Se_2$ corresponding to y = 0.4 (superstructure **I**) and y = 0.5 (superstructure **II**). Vacant Fe columns are shown as squares.

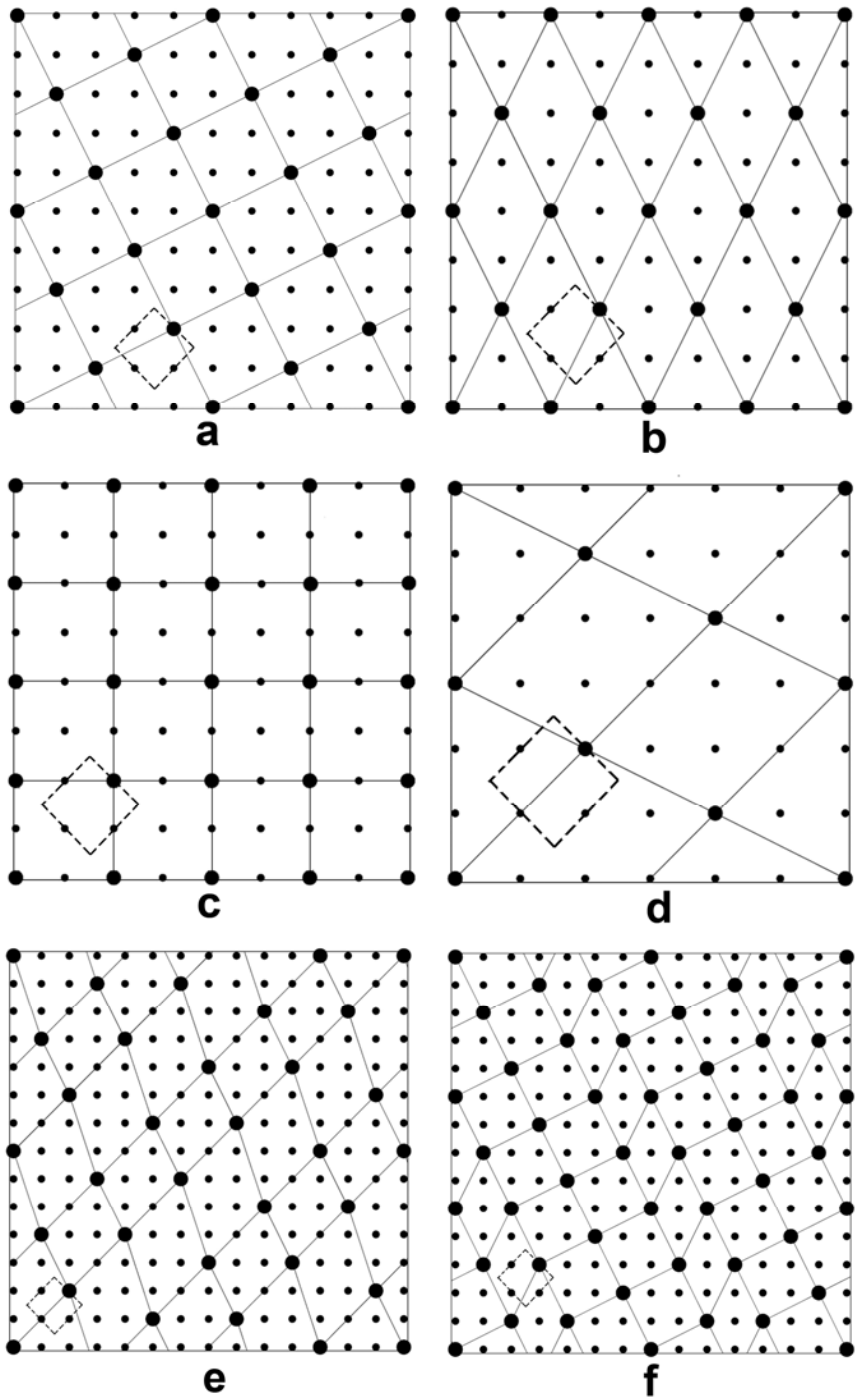

**Figure 8.** Possible Fe-vacancy ordering patterns in $K_x(Fe,Co)_{2-y}Se_2$ for different y values: a) uniform tetragonal y = 2/5 pattern (superstructure **I**); b) uniform orthorhombic y = 1/2 pattern (superstructure **II**); c) hypothetical unobserved non-uniform tetragonal y = 1/2 pattern; d) uniform oblique y = 1/3 pattern; e) uniform stripe y = 2/7 pattern; f) uniform stripe y = 3/7 pattern. The Fe atoms and vacancies are shown as small and large circles, respectively. The $ThCr_2Si_2$-type subcell is marked with dashed square.